\documentclass{jpconf}
\usepackage{graphicx}
\usepackage{amsmath}
\usepackage{amssymb}
\usepackage{lmodern}

\begin{document}
\title{On properties of the Wang--Landau algorithm}
\author{L.\,N.~Shchur$^{1,2,3}$}
\address{$^1$ Landau Institute for Theoretical Physics, 142432 Chernogolovka, Russia}
\address{$^2$ Science Center in Chernogolovka, 142432 Chernogolovka, Russia}
\address{$^3$ National Research University Higher School of Economics, 101000 Moscow, Russia}

\begin{abstract}
We review recent advances in the analysis of the Wang--Landau algorithm,
which is designed  for the direct Monte Carlo estimation of the density of
states (DOS). In the case of a discrete energy spectrum, we present an
approach based on introducing the transition matrix in the energy space
(TMES). The TMES fully describes a random walk in the energy space biased
with the Wang--Landau probability. Properties of the TMES can explain some
features of the Wang--Landau algorithm, for example, the flatness of the
histogram. We show that the Wang--Landau probability with the true DOS
generates a Markov process in the energy space and the inverse spectral gap
of the TMES can estimate the mixing time of this Markov process. We argue
that an efficient implementation of the Wang--Landau algorithm consists of
two simulation stages: the original Wang--Landau procedure for the first
stage and a $1/t$ modification for the second stage. The mixing time
determines the characteristic time for convergence to the true DOS in the
second simulation stage. The parameter of the convergence of the estimated
DOS to the true DOS is the difference of the largest TMES eigenvalue from
unity. The characteristic time of the first stage is the tunneling time,
i.e., the time needed for the system to visit all energy levels.
\end{abstract}

\section{Introduction}

The history of Monte Carlo simulations begins in the early years of
computing. The method was developed in one of the most cited
papers in computational physics and chemistry~\cite{Metropolis}, where both
the Monte Carlo method and the molecular dynamics approach were introduced.
Many algorithms for Monte Carlo simulations have since been developed. In
statistical mechanics, different Monte Carlo algorithms are based on
different partition function representations.

A general partition function representation is
\begin{equation}
Z=\sum_{\left\{i\right\}} e^{-E_i/k^{}_{\mathrm{B}}T},
\label{partition-general}
\end{equation}
where $E_i$ is the energy of the configuration $\left\{i\right\}$ of the
system in the phase space, $T$ is the temperature, and $k^{}_{\mathrm{B}}$ is the
Boltzmann constant. Representation~(\ref{partition-general}) can be
associated, for example, with the Metropolis algorithm~\cite{Metropolis},
which is based on the local updates taking energy changes $\Delta E_{kl}=
E_l-E_k$ with moves in the phase space from state $\left\{k\right\}$ to
state $\left\{l\right\}$ into account. The probability of accepting the
move is given by the Metropolis probabilities
\begin{equation}
P=\min \left[ 1, e^{-\Delta E_{kl}/k^{}_{\mathrm{B}}T} \right].
\label{Metropolis-prob}
\end{equation}
Modifications of the Metropolis algorithm are based on variations of local
updates, as well as more complicated algorithms, which in any event use
local update rules and probabilities (see book~\cite{Binder-Landau} for
references). All of them use partition function~(\ref{partition-general}) as
the basic representation.

A more sophisticated algorithm named the MuMC algorithm (multi-canonical
Monte Carlo) is based on another representation of the partition
function~\cite{Berg-Neuhaus-1,Berg-Neuhaus-2},
\begin{equation}
Z=\sum_{k=1}^{N_E} g(E_k) W(E_k),
\label{partition-MuMC}
\end{equation}
where $N_E$ is the number of energy levels, $g(E_k)$ is the number of states
with the energy $E_k$, and $W(E )$ is an a priori unknown weight function,
which is estimated in the simulations. The condition to stop is that all
$g(E_k) W(E_k)$ are approximately equal to each other. It is successfully
used in polymer simulations~\cite{Janke-Paul}.

Another famous example is based on a cluster representation~\cite{Fortuin-Kasteleyan}
of the Potts model,
\begin{equation}
Z=\sum_{\mathrm{bonds}} p^{b}(1-p)^n q^{N_c},
\label{partition-SW}
\end{equation}
where $p=1-\exp({-J/k^{}_{\mathrm{B}}T})$, $J$ is a constant measured in
energy units, $b$ is the number of bonds connecting spins, $n$ is the number
of bonds not connecting spins, $q$ is the number of spin components, and
$N_c$ is the number of clusters in the configuration. The cluster
representation leads to several efficient algorithms for simulating the
Potts model; the two most famous are the Swendsen--Wang cluster
algorithm~\cite{Swendsen-Wang} and the Wolf one-cluster algorithm~\cite{U-Wolf}.

The abovementioned algorithms are extensively used in simulations. We do not
discuss the pros and cons of the algorithms and refer our readers to the
book~\cite{Binder-Landau}.

A common property of all the mentioned algorithms is that simulations depend
explicitly on the temperature $T$. The algorithm that we review and analyze
here is quite different: simulations are independent of the temperature.

The Wang--Landau (WL) algorithm is a way to estimate the density of states
(DOS) directly~\cite{Wang-Landau,Wang-Landau-PRE}. It can be applied to any
system with a defined partition function. Its idea is based on representing
the partition function as a sum over the energy levels (somewhat similar to
representations used for the MuMC algorithm),
\begin{equation}
Z=\sum_{k=1}^{N_E}g(E_k)e^{-E_k/k^{}_{\mathrm{B}}T}.
\label{partition-function}
\end{equation}
It has proved to work quite well for many systems, with applications in
different areas of statistical physics and mechanics. There are more than
1500 papers on the application of the algorithm and its improvements. At the
same time, not all properties of the algorithm are well understood. For
example, it is unclear how auxiliary histogram flatness is related to the
convergence of the algorithm and how accuracy of the DOS estimation evolves
with the simulation time.

An approach based on constructing the transition matrix in the energy space
(TMES) was recently proposed~\cite{BFS-TM}. It answers the above questions
and can be used for a deeper analysis of the algorithm and for modifications
of the WL algorithm. We present a summary of the approach to the analysis of
the WL algorithm. We regard the WL algorithm as a random walk in the energy
space as soon as the WL probabilities of moves between energies are based
purely on the ratios of the corresponding densities $g(E)$ of the energy
states. We explain some properties of the WL algorithm, including the
flatness criteria and the characteristic time scales, and discuss the
control parameter for estimating the accuracy of the DOS estimate and for
the convergence of the DOS to the true values.

In section~\ref{sec-WL}, we briefly review the WL algorithm. In
section~\ref{sec-TMES}, we introduce the TMES. In section~\ref{sec-time}, we
define and discuss the characteristics time of the WL algorithm. The
discussion in section~\ref{sec-disc} finishes our review.

\section{Wang--Landau algorithm}
\label{sec-WL}

We briefly describe the WL algorithm here. We take a configuration of the
system, compute the energy value $E_k$, randomly choose an update to a new
configuration with the energy $E_m$, and accept this configuration with the
WL probability
\begin{equation}
{\widetilde P_{\mathrm{wl}}}=\min\left[1,\frac{\tilde g(E_k)}{\tilde g(E_m)}\right],
\label{WLP-approx}
\end{equation}
where $\tilde g(E)$ is the current DOS approximation. The approximation is
obtained recursively by multiplying $\tilde g(E_m)$ by a factor $f$ at each
step of the random walk in the energy space. Each time that the auxiliary
histogram $H(E)$ becomes sufficiently flat, the parameter $f$ is modified by
taking the square root, $f:=\sqrt{f}$. Each histogram value $H(E_m)$ counts
the number of moves to the energy level $E_m$. The histogram is filled with
zeros after each modification of the refinement parameter $f$. It is
convenient to work with the logarithms of the values, $S(E_k):=\ln\tilde g(E_k)$
and $F:=\ln f$, and to replace the multiplication $\tilde g(E_m):=f\cdot\tilde g(E_m)$
with the addition $S(E_m):=S(E_m)+F$. At the end of the simulation, the
algorithm provides only a relative DOS. Either the total number of states or
the number of ground states can be used to determine the absolute DOS.

\section{Transition matrix in the energy space}
\label{sec-TMES}

The central idea of the WL algorithms is the WL probability, which generates
a random walk in the energy space. We first argue that random walk in the
energy space with the WL probability calculated using the true DOS is a
Markov chain and satisfies the detailed balance condition~\cite{BFS-TM}. The
WL probability in this case is given by
\begin{equation}
P_{\mathrm{wl}}=\min\left[1,\frac{g(E_k)}{g(E_m)}\right],
\label{WLP-true}
\end{equation}
where $g(E)$ is the true DOS (cf.~WL probability~(\ref{WLP-approx})).

It is useful to consider a TMES whose elements show the frequency of
transitions between energy levels during the WL random walk in the energy
space~\cite{BFS-TM}. Its elements are influenced by both the random process
of choosing a new configurational state and the WL probability of accepting
the new energy,
\begin{equation}
T(E_k,E_m)=\min\left(1,\frac{g(E_k)}{g(E_m)}\right)P(E_k,E_m),
\label{TMES-true}
\end{equation}
which represents nondiagonal elements of the TMES of the WL random walk in
the energy space. Here, $P(E_k,E_m)$ is the probability of one step of the
random walk to move from a configuration with the energy $E_k$ to any
configuration with the energy $E_m$.

The random walk in the configuration space is a Markov chain. Its invariant
distribution is uniform, i.e., the probabilities of all states of the
physical system are equal to each other. For any pair of configurations
$\Omega_A$ and $\Omega^{}_{\mathrm{B}}$, the probability of an update from $\Omega_A$ to
$\Omega^{}_{\mathrm{B}}$ is equal to the probability of an update from $\Omega^{}_{\mathrm{B}}$ to
$\Omega_A$. Hence, the detailed balance condition is satisfied. Therefore,
\begin{equation}
g(E_k)P(E_k,E_m)=g(E_m)P(E_m,E_k),
\label{simplebalance}
\end{equation}
where $g(E)$ is the true DOS. It follows from (\ref{TMES-true}) and
(\ref{simplebalance}) that
\begin{equation}
T(E_k,E_m)=T(E_m,E_k).
\label{TMES-symm}
\end{equation}
Therefore, the TMES of the WL random walk on the true DOS is a symmetric
matrix. Because the matrix is both symmetric and right stochastic, it is
also left stochastic. Therefore, the sum over rows (columns) is equal to
unity, and the rates of visiting of all energy levels are equal to each
other. The corresponding auxiliary histogram of energy visits in the WL
algorithm is indeed flat! (See the discussion in section~\ref{sec-disc}.)

We therefore observe~\cite{BFS-TM} that the flatness of the histogram in the
WL algorithm is just the proximity of the corresponding random walk in the
energy space to the Markov process. The closer the estimated DOS is to the
true DOS, the closer the TMES is to the fully stochastic matrix. This
property can be used to connect control of DOS convergence to the property
of TMES convergence to the stochastic matrix.

\subsection{Exact TMES for 1D Ising model}

It is instructive to seek an exact solution of TMES elements. This can be
easily done for the 1D Ising model~\cite{BFS-TM}. In the case of a chain of
$L$ spins with periodic boundary conditions, the probability to change the
energy from $E_k$ to $E_m$ in a WL random move is
\begin{equation}
T(E_k,E_m)=\min\left[1,\frac{g(E_k)}{g(E_m)}\right]
\sum_{i=0}^{2k}\frac{N_iQ_i^{E_k\to E_m}}{g(E_k)},
\label{el-mat-1d}
\end{equation}
where $k\ne m$. Here, $k$ is the number of couples of domains walls in the
configuration, which determines the energy level $E_k=-\sum_{j=1}^{L}
\sigma_j\sigma_{j+1}=-L+4k$, $N_i(k,L)$ is the number of configurations
where $i$ domains consist of only one spin and $2k{-}i$ domains consist of
more than one spin, and $Q_i^{E_k\to E_m}(L)$ is the probability that a
single spin flip moves system to the energy $E_m$ from such configurations.

The structure of expression (\ref{el-mat-1d}) shows that TMES elements are
composed of three probabilities: the probability of the particular
configuration, the probability that the move in the configuration space
changes the system energy to some value, and the WL probability
$P_{\mathrm{wl}}$ to accept the proposed move between energy levels. It
reflects the inside of the WL algorithm and explicitly shows how the
combination of choosing the site in the configuration space and of the
corresponding change in the phase space generates a random walk in the
energy space.

Occupations of energy levels of the chain are expressed in terms of
binomial coefficients as $g(E_k)=2C_L^{2k}$ because there are exactly
$C_L^{2k}$ ways to arrange $2k$ domain walls. Therefore, partition function
(\ref{partition-function}) is
\begin{equation}
Z_L=2\sum_{k=0}^{L/2} C_L^{2k} e^{(L-4k)/(k^{}_{\mathrm{B}}T)}
\label{Z_L}
\end{equation}
and the TMES for the 1D Ising model is
\begin{equation}
T(E_k,E_{k+1})=T(E_{k+1},E_k)=\frac{C^{2k}_{L-2}}{\max\left(C^{2k}_L,C^{2k+2}_L\right)}.
\label{tmatrix1d}
\end{equation}
The TMES is indeed symmetric and stochastic.

\subsection{Estimation of TMES elements during simulation}

We estimate the TMES elements in simulations as follows~\cite{BFS-TM}. The
auxiliary matrix $U(E_k,E_m)$ is initially filled with zeros. The element
$U(E_k,E_m)$ is increased by unity after every WL move from a configuration
with the energy $E_k$ to a configuration with the energy $E_m$. During the
simulations, we compute the normalized matrix
\begin{equation}
\widetilde{T}(E_k,E_m)=U(E_k,E_m)/\widetilde{H},
\end{equation}
where
\begin{equation}
\widetilde{H}=\sum_{k,m}U(E_k,E_m)/N_E.
\end{equation}
The obtained matrix $\widetilde{T}$ is expected to approach the stochastic
matrix $T$ in the final stage of correct calculation.

\section{Two characteristic times in WL algorithm}
\label{sec-time}

Results of simulations~\cite{BFS-TM,FS-mix} demonstrate that there are two
time scales in the WL simulations (although this point is not discussed in
the texts). It is well known that the WL algorithm drives the estimated DOS
to the vicinity of the true DOS at the beginning of simulations, at least
for systems with a discrete energy spectrum and a not very complicated
energy landscape. This is a positive feature of the WL algorithm, although
still unexplained. The time scale of this stage can be associated with the
tunneling time. The second stage is the stage of convergence of the
estimated DOS to the true DOS. In this section, we argue that the time scale
of the second stage is given by the inverse value of the spectral gap of the
TMES.

\subsection{Tunneling time}

Is it known from the beginning of WL simulations~\cite{Landau-talk} that the
finite value of the desired histogram flatness leads to saturation of the
estimates, and saturation was found for systems with a known DOS. The first
stage of the WL algorithm drives the estimated DOS close to the true DOS in
the first simulation stage, although not very close. Analysis of Figures~1
and 3--5 in paper~\cite{BFS-TM} shows that there is a characteristic time of
the first stage; we suggest that it is the tunneling time of the random walk
in the energy space. There are several definitions leading to the same
scaling behavior of the tunneling time with system size (more precisely,
with the number of energy levels), $T_{\mathrm{t}}\propto L^z$. A useful practical
definition is that it is the time needed for the WL random walk to reach the
highest energy level starting from the lowest one~\cite{Dayal2004}.

The modification of the WL algorithm called the $1/t$-WL algorithm~\cite{BP-1,BP-2}
defines the time when the $1/t$ regime of the WL algorithm starts as the
time when all energy levels have been visited at least once. Clearly, it is
practically the same as the tunneling time.

The stochastic approximation Monte Carlo (SAMC) algorithm~\cite{SAMC-1,SAMC-2}
does specify how to choose the time $t_0$ when the $1/t$ regime starts. Our
preliminary analysis shows that the optimal choice of $t_0$ is indeed the
tunneling time.

There is a natural lower limit for the exponent $z$. It can be understood in
terms of the random walk. We first consider the classical problem of the 1D
unbiased random walk: the position $i$ of the random walk changes with equal
probability to $i-1$ or $i+1$. The point~0 is the reflecting point of the
random walk, and the point $N$ is the absorbing point. The first-passage
time of the random walk is known to scale as $N^2$. The number of levels
$N_E$ for the 2D Ising model scales as $L^2$, and the number of random walk
jumps necessary to cross all energy spectrum hence scales as $L^4$.
Therefore, $z=4$ is the minimum value for any type of random walk in the
energy space of the 2D Ising model.

For a random walk with bias, which is the WL probability $P_{\mathrm{wl}}$,
the growth of the tunneling time should be faster, $z>4$. There is a worse
message, which can be learned from the properties of a random walk in a
random potential. It states that localization of the random walker occurs in
one dimension~\cite{Sinai}. This probably explains the difficulties in
applying the WL algorithm to systems with complex energy landscapes: the
random walker can be temporarily trapped near a global minimum of the
probability profile in the simulated DOS.

Positions of the vertical dashed line in Figures~1 and 3--5 in
paper~\cite{BFS-TM} can be associated with the tunneling time. During the
first simulation stage, the effective tunneling time exponent grows from
$z=4$ to a value that depends on the investigated system. Reported
estimated exponents for the scaling of the tunneling time are $z=4.743(7)$
for the 2D Ising model on a square lattice of linear size $L$ and
approximately $z=5.7$ for the fully frustrated 2D Ising model~\cite{Dayal2004}.

\subsection{Mixing time}

The spectral gap $G$ of the TMES determines the mixing time of the Markov
chain~\cite{Boyd2004}:
\begin{align}
&T_{\mathrm{m}}=1/G,
\label{tmix}
\\
&G=\lambda_1-\lambda_2,
\label{gap}
\end{align}
where $\lambda_1$ and $\lambda_2$ are the two largest eigenvalues of the TMES.

In the second phase of the $1/t$-WL (or SAMC) algorithm, the DOS is very
close to the true DOS, and the characteristic time of convergence is
therefore just the mixing time $T_{\mathrm{m}}$ of the Markov chain.

The exact solution for the 1D Ising model~\cite{BFS-TM} can be used to
construct the TMES for varying lattice sizes, and fitting to the inverse
spectral gap yields an estimate~\cite{FS-mix} for the mixing time exponent
$T_{\mathrm{m}}\propto L^{2.19}$. Numerically estimating the mixing time for the 2D
Ising model~\cite{FS-mix} leads to the result $T_{\mathrm{m}}\propto L^{4.37(2)}$.

\section{Discussion}
\label{sec-disc}

Using our approach, we can calculate the normalized histogram ${\cal H}=
H(E_m)/\sum_{m}H(E_m)$ from the TMES ${\cal H}=\sum_{k}\widetilde{T}(E_k,E_m)$.
It is obvious that the histogram flatness condition is equivalent to the
property that the matrix $\widetilde{T}$ is close to the stochastic matrix. The
histogram flatness in the second simulation stage is closely related to the
stochastic properties of the TMES because the estimated DOS is very close to
the true DOS.

We have argued that the time $t_{\mathrm{c}}$ in the $1/t$-WL algorithm~\cite{BP-1,BP-2}
and the time $t_0$ in the SAMC algorithm~\cite{SAMC-1,SAMC-2} is of the
order of the tunneling time~\cite{Dayal2004}. Therefore, in the first
simulation stage, the optimal choice is the original WL algorithm, which
drives the estimated DOS close to the true DOS, at least for systems with
a not very complicated free energy landscape. In the second simulation
stage, the $1/t$-WL algorithm drives the estimated DOS to the true DOS, and
the characteristic time is the mixing time $T_{\mathrm{m}}$. The accuracy of estimating
the DOS is given by the deviation of the largest eigenvalue of the estimated
TMES from unity~\cite{BFS-TM}.

Future research should be done to understand properties of the WL algorithm
when applied to more complicated systems and to systems with a continuous
energy spectrum. In systems with a complex energy landscape, the variation
of the transition matrix elements can be huge, and this can lead to a large
increase of the tunneling time. It follows from the theory of a random walk
in a random potential~\cite{Sinai} that the random walker can be localized
in an extremal valley of the potential. It is important to understand under
which conditions this can happen with the WL algorithm.

Partitioning a continuous energy spectrum into bins (done in simulations)
allows using the TMES approach to analyze simulations. It would be
instructive to understand how the bin size influences TMES properties, from
which we might understand some properties of simulations, the behavior of
the control parameter, and also the tunneling and mixing time values.

\medskip

This work was supported by grant 14-21-00158 from the Russian Science Foundation.


\end{document}